\documentstyle[12pt]{article}
\newcommand{\bce}{\begin{center}}
\newcommand{\ece}{\end{center}}
\newcommand{\beq}{\begin{equation}}
\newcommand{\eeq}{\end{equation}}
\newcommand{\bea}{\vspace{0.25cm}\begin{eqnarray}}
\newcommand{\eea}{\end{eqnarray}}

\newcommand{\ba}{\begin{array}}
\newcommand{\ea}{\end{array}}

\def\lsim{\mathrel{\rlap{\lower4pt\hbox{\hskip1pt$\sim$}}
    \raise1pt\hbox{$<$}}}         
\def\gsim{\mathrel{\rlap{\lower4pt\hbox{\hskip1pt$\sim$}}
    \raise1pt\hbox{$>$}}}         

\def\lsim{\mathrel{\rlap{\lower4pt\hbox{\hskip1pt$\sim$}}
    \raise1pt\hbox{$<$}}}         
\def\gsim{\mathrel{\rlap{\lower4pt\hbox{\hskip1pt$\sim$}}
    \raise1pt\hbox{$>$}}}         

  \advance\oddsidemargin  by -1in
  \advance\evensidemargin by -1in
  \advance\topmargin      by -0.5in
\def\lsim{\mathrel{\rlap{\lower4pt\hbox{\hskip1pt$\sim$}}
    \raise1pt\hbox{$<$}}}         
\def\gsim{\mathrel{\rlap{\lower4pt\hbox{\hskip1pt$\sim$}}
    \raise1pt\hbox{$>$}}}         

\def\beq{\begin{equation}}
\def\endeq{\end{equation}}
\def\arr{\begin{eqnarray}}
\def\endarr{\end{eqnarray}}
\makeindex
 
\begin{document}
 
\begin{flushright} 
{\sl FZ IKP(TH)-2001-25}
\end{flushright}
\begin{center}
{\Large \bf
Coherent Coulomb 
excitation of relativistic nuclei in aligned  crystal targets
 \vspace{1.0cm}}
 
{\large \bf
V.R.Zoller\medskip\\ }
{\sl  Institute for Theoretical and Experimental Physics \\
117218 Moscow Russia
{\footnote {\rm zoller@heron.itep.ru}}\vspace{1cm}\\}
{\bf           Abstract}
\end{center}
We study coherent   Coulomb excitation of 
ultrarelativistic nuclei passing through the aligned crystal target.
We develop multiple scattering theory description of 
this process  which consistently incorporates  both the 
specific resonant properties of particle-crystal interactions and the
shadowing  effect 
typical of the diffractive scattering.
We emphasise that the effect of quantum mechanical diffraction
makes the physics of ultrarelativistic nuclear excitations entirely
different from the physics of non-relativistic atomic excitations
experimentally studied so far.
 It is found that
at small transverse momenta $q_{\perp}$ 
the shadowing effect drastically changes
the  dependence of coherent amplitudes on the 
crystal thickness $L$, 
from the widely discussed  
growth $\propto  L$ typical of the
Born approximation to
the inverse thickness  attenuation law.  
At relatively large $q_{\perp}$ no attenuation effect is found but
 the coherency condition is shown to   put stringent constrain on the growth 
of the transition rate with growing  $L$.

\newpage

\section{} 
There were several proposals  of  experiments
 on 
the coherent Coulomb excitation of relativistic nuclei in crystal
 target \cite{PROP1, PROP2,FUSINA,PROP3}
\beq
A\gamma\to A^*. \label{eq:ATOA}
\endeq 
The central idea of \cite{PROP1,PROP2,FUSINA,PROP3} is that
in a nucleus (the ground state $ |0\rangle$
 and the excited state $|1\rangle$)   under periodical perturbation
$V\sin{\nu t}$ with the frequency $\nu$ equal to
the level splitting  $\Delta E=E_1-E_0$ there develop quantum beats with
the oscillation frequency $\omega= \langle 1|V|0\rangle$. 
 If the perturbation $V$ is weak, then for $\omega t\ll 1$
the $|0\rangle\to |1\rangle$  transition probability $P(t)$ 
increases rapidly with the time
$
P(t)\,\propto\,\omega^2t^2\,.$          
 If one could 
subject nuclei to  a high frequency field, 
$\nu\simeq\Delta E $,
 the rates of transition can be enhanced substantially.
The high  monochromaticity
is an evident condition to sustain the fast  growth of $P(t)$ over large 
time scale.  
It has been  
  suggested in \cite{PROP1,PROP2,FUSINA,PROP3} 
that all of the  above requirements are met
 best 
in Coulomb interaction of a  
high-energy nucleus propagating in a crystal along the crystallographic axis.
 Here the r\^ole of "time" is  played by the crystal thickness $ L$. 
 For ultrarelativistic particles $\nu$ is enhanced due to
 the Doppler shift, $\nu^{\prime} = \gamma\nu$, where $\gamma$ is the Lorentz
factor. It is the Lorentz factor which can boost $\nu$ to the hundreds
 KeV's range.

In  \cite{PROP1,PROP2,FUSINA,PROP3} 
the Coulomb field of a crystal was evaluated in the Weizs\"aker-Williams 
approximation,  and then applied to the calculation of 
the transition amplitude in the plane wave  Born approximation.
It was claimed \cite{PROP1,PROP2,PROP3} that the law  
$
P(N)\,\propto\,N^2\,,
$ 
holds  up to the crystal thicknesses  $N=L/a\sim 10^4-10^5 $  
(hereafter $a$ stands for the lattice spacing). 
However, in a consistent treatment of coherent $A\to A^*$ transitions 
one needs to include
distortions due to the initial 
state Coulomb interactions (ISI) of
 the nucleus $A$ and final state Coulomb interactions (FSI) 
of the excited nucleus $A^*$.

In this paper we show that at small transverse momenta the effect of
these interactions (multiple diffractive scattering off atomic row)
is reminiscent
the  well known  Glauber-Gribov shadowing  
effect \cite{GLAU,GRIB} and entails 
 an  attenuation of the coherent excitation amplitudes 
with growing crystal thickness.
The early discussion of  shadowing  in the total cross section
of  elastic high-energy  particle-crystal scattering  
is found in \cite{KALASH}.

Here we emphasise that  it is the diffraction phenomenon
which makes 
the  physics  of ultrarelativistic nuclear  excitations with
$\gamma \sim 10$ entirely  different from the
physics of coherent  excitation of atoms \cite{OKOR,EXP}, where typically 
 $\gamma\ll 1$.

It should be noticed, however,  that in contrast to 
the high-energy hadronic scattering which is described by
 predominantly imaginary amplitudes, the shadowing in  Coulomb 
scattering does not
imply a simple "absorption", but a redistribution of scattered nuclear waves 
in the  phase space. Therefore the depletion of the domain of small-$q_{\perp}$
means the enhancement of the large-$q_{\perp}$ region.  
At relatively large transverse momenta
 no shadowing is found but
 the nuclear  waves scattered by different  atoms 
get out of tune easily. Here the structure factor of crystal
 puts stringent constrain  on 
the coherent excitation rate which at the  projectile energies available
 would not exhibit  any sizable enhancement effect.

\section{}

Consider the  small-angle Coulomb scattering of ultrarelativistic nucleus 
(the mass number $A$, the  charge $Z_1$ and  the four-momentum $p$)
  moving along a crystal
 axis.  The projectile-nucleus undergoes a correlated series of 
 soft  collisions
which give rise to diagonal ($A\to A$, $A^*\to A^*$) and
 off-diagonal ($A\to A^*$, $A^*\to A$)
 transitions.  
The interatomic distances, $a\sim 3-5\AA$,   are large,
compared to the  
 Thomas-Fermi screening radius $r\simeq 0.468Z_2^{-1/3}\AA$, 
 where $Z_2$ is the atomic number of the target atom and $\alpha=1/137$. 
The  amplitude of thermal vibrations of the lattice 
$\langle{ u}^2\rangle^{1/2}$
estimated from the Debye approximation
equals to $\langle{ u}^2\rangle^{1/2}\sim (0.05-0.1)\, \AA$
for most commonly studied crystals at room temperature \cite{GEM}.
The relevant impact parameters, $b$, 
satisfy as we shall see $b\ll a$.
 This implies that the amplitudes of 
scattering by different atomic strings  parallel to a given  crystallographic 
axis are incoherent. 

 In \cite{FUSINA,PROP1,PROP2} it has been proposed to study
the electric dipole transition in $^{19}F$,
  the excitation of the state $|J^{\pi}={1/2}^-\rangle$ 
from the ground state $|1/2^+\rangle$.
 Let us accept this proposal and consider
 the phenomenological   matrix element of the transition $1/2^+\to 1/2^-$
\beq
{\cal M}=d\bar u(p^{\prime}) 
\gamma_5 
\sigma_{\mu\nu} u(p)q^{\nu}\varepsilon^{\mu}, \label{eq:M}
\endeq
where both $u(p^{\prime})$ and $ u(p)$ are bispinors of initial
 and final states of the
projectile-nucleus. By definition, 
$\sigma_{\mu\nu}={1\over 2}[\gamma_\mu,\gamma_\nu]$, $d$ is the transition
 dipole moment,
 $p^{\prime}=p+q$, ${\bf q}=({\bf q_{\perp}},\kappa)$,
${\bf q_{\perp}}=(q_{\perp}\cos\phi,q_{\perp}\sin\phi)$ 
 and $\varepsilon$
 is the photon polarisation vector.
Then after a series of quite standard high-energy approximations 
one readily finds the helicity-flip Born amplitude $t_B({\bf q}_{\perp})$
 of the transition $1/2^+\to 1/2^-$
in  the 
nucleus-atom collision 
\beq
{t}_B({\bf q_{\perp}})=
\sqrt{\alpha}dZ_2{q_{\perp}e^{i\phi}\over q^2_{\perp}+\lambda^2 }
\,,               
    \label{eq:GQ}
\endeq 
where  $\lambda^2=\mu^2+\kappa^2$ and $\mu=r^{-1}$. 
Denoted by $\kappa$ is the longitudinal momentum transfer
\beq
\kappa={2M\Delta E+q^2_{\perp}\over 2p},
\label{eq:kappa}
\eeq
which determines the coherency length $l_c\sim \kappa^{-1}$.
In eq.(\ref{eq:kappa}) $M$ is  the mass of  
projectile-nucleus. The excitation energy is $\Delta E\simeq 110$ KeV.
Normalisation of amplitudes is such that
\beq
{d\sigma\over dq_{\perp}^2}=|t({\bf q}_{\perp})|^2.
\label{eq:DIFSEC}
\eeq
 The integral $\sigma_B=\int dq_{\perp}^2 |t_B({\bf q}_{\perp})|^2$
diverges at large $q_{\perp}$ and need be regularised. The natural regulator is
 the inverse amplitude of thermal vibrations of the  lattice.

Making use of the impact parameter representation simplifies the summation
of diagrams of multiple Coulomb scattering.
Then, in the eikonal approximation 
 the full nucleus-atom amplitude to all orders in $\alpha Z_1Z_2$
 reads
\beq
t({\bf q_{\perp}})
=\sqrt{\alpha}{dZ_2\lambda}e^{i\phi}\int bdbJ_1(q_{\perp}b)K_1(\lambda b)
\exp[i\chi(b)]\,,
                   \label{}
\endeq  
where $J_1(x)$ and $K_{0,1}(x)$ are the  Bessel functions and  
the screened Coulomb phase shift function is  
\beq
\chi(b)  =-\beta K_0(\mu b),
 \label{eq:FBORN} 
\endeq
where $\beta=2\alpha Z_1Z_2$.

The only phenomenological parameter of eq.(\ref{eq:M}), 
 dipole moment of the $1/2^+\to 1/2^-$ transition, denoted by  $d$,
 can be  
determined from the width $\Gamma$ of the $110$ KeV   level $^{19}F(1/2^-)$
which is
$
\Gamma={d^2\Delta E^3/ \pi}+{\cal O}\left({\Delta E/ M}\right).
$
Then the measured life-time 
$\tau=\Gamma^{-1}=(0.853\pm 0.010)\times 10^{-9}$ sec \cite{Ajzen}
 yields
 $d\simeq 4.3\times 10^{-8}$ KeV$^{-1}$.
 Two useful conclusions   can be 
 drawn immediately.
First, because of large value of $\tau$
the decay of excited state inside the target can be safely neglected. 
Second, due to the smallness of $d$, the excitation amplitude
is much smaller than the elastic Coulomb 
amplitude for all $q_{\perp}$ up to $q_{\perp}\sim \sqrt{\alpha}Z_1/d$ and can
 be considered as a perturbation. Thus the multi-channel problem reduces to
the one-channel one.

Then, in the static lattice approximation 
 the evaluation of the  full transition amplitude on a string 
 of
 $N$ identical atoms  reads \cite{JETPLET}
\beq
{T}({\bf q}_{\perp})=\sqrt{\alpha}
d Z_2
\lambda S(q_{\perp})e^{i\phi}I(q_{\perp}),                                             
\label{eq:TSIGMA} 
\endeq 
where
\beq
I(q_{\perp})= \int bdb\,J_1(q_{\perp}b)K_1(\lambda b)\exp[-i\beta N K_0(\mu b)]
\label{eq:I} 
\endeq
In eq.(\ref{eq:TSIGMA}) $S(q_{\perp})$ is the structure factor of the lattice 
\beq
S(q_{\perp})={{\sin(\kappa Na/2)}\over{\sin(\kappa a/2)}}\,,
                                        \label{eq:SL}
\endeq
and $\kappa$ is given by eq.(\ref{eq:kappa}).

Split  integration over $b$ in eq.(\ref{eq:I}) into the two domains:
$\mu^{-1}\lsim b\lsim a$ and $0<b<\mu^{-1}$.
In the former domain it suffices to use 
the asymptotics $K_{0,1}(x)\sim \exp(-x)$ which upon the
slight readjustment of screening parameters, 
$\mu\to\mu^{\prime}=\mu(1+{1\over 2}\log(2\xi/\pi))$ and
$\lambda\to\lambda^{\prime}=\mu(1+{1\over 2}\log(2\xi/\pi))$
at the relevant  $b\sim\mu^{-1}\xi$
proves numerically
very accurate.  Hereafter, 
$\xi=\log(\beta N/\Delta)$ and
 $\Delta=\lambda/\mu=\sqrt{1+\kappa^2/\mu^2}$.
Then, the  steepest descent from the  saddle-point  at
$$ b_0=\mu^{-1}(\xi+i\pi/2),$$ 
 yields for $\xi^2\gg \mu/\lambda$ and  $q_{\perp}\lsim\mu \xi^{-1}$  
\bea
I(q_{\perp})\simeq {q_{\perp}\over \mu^3}\sqrt{2\pi\over \Delta}e^{-\Delta}
\log^2\left({i\beta N}\over \Delta\right)
\left(\Delta\over i\beta N\right)^{\Delta} \\ \nonumber
\propto \log^2N\left({1\over N}\right)^{\Delta} \,.
                                                 \label{eq:TESTIM}
\eea 
Consequently, for small transverse momenta $q_{\perp}\lsim\mu\xi^{-1}$ as soon as 
$\xi \gg 1$  which holds
for all practical purposes  one
 has the attenuation of the coherent excitation amplitude.

For higher transverse momenta $ q_{\perp}\gg \mu\xi^{-1}$ making use of  
 the stationary phase approximation with the above reservation about
the substitution $\mu\to\mu^{\prime}$ and $\lambda\to\lambda^{\prime}$
 yields
\bea
I(q_{\perp})\simeq {\sqrt{\eta}\over \mu q_{\perp}}
\exp(-\Delta \eta) 
\exp\left[-i{q_{\perp}\over \mu}(\eta+1)\right]\\ \nonumber
\propto \sqrt{\log N}\left({1\over N}\right)^{\Delta}
\,,
                                                 \label{eq:HQ}
\eea
 where $\eta=\log(\mu\beta N/q_{\perp}) \gg 1$.
As before we find no enhancement but  the attenuation of 
the coherent transition amplitude.
Indeed, let the projectile momentum satisfies 
 the resonance condition \cite{PROP1,PROP2,FUSINA,PROP3} 
\beq
 {M\Delta E\over p}={2\pi n\over a}\,,\,\,\,n=0,\,1,\, 2...\,.\label{eq:KAPRES}
\endeq
In \cite{PROP1,PROP2,PROP3} it has been suggested to look for the
 coherent  transitions in the $W$-crystal  at $n=3$.
This regime  corresponds to $\gamma\simeq 10$ and  $\Delta
=\lambda/\mu\simeq\sqrt{1+4\pi^2 n^2/(a\mu)^2}\simeq 1.2$. 
Still higher $n$ discussed in \cite{PROP1,PROP2,FUSINA,PROP3}
 would correspond
to higher momentum transfers and 
to  much stronger suppression of the coherent excitation  amplitude.

Now consider the contribution to $I(q_{\perp})$ from 
 the second domain, $0<b<\mu^{-1}\equiv r$.
The region of small impact parameters is
 affected by the lattice thermal vibrations
which are known to suppress the coherent amplitude.
Let us make use of  the fact that for some commonly used  crystals
at a room temperature the root-mean-square one-dimensional 
 displacement $u$  is such that $u\ll r$. For example,
for the diamond  lattice suggested as a target in \cite{FUSINA}
 $u/r\simeq 0.16$  \cite{GEM}. 
The integral in $r.h.s.$ of 
eq.(\ref{eq:I}) reads
\bea                                             
I(q_{\perp})=\int^{r}_0 bdb\,J_1(q_{\perp}b)K_1(\lambda b)
\exp[-i\beta N K_0(\mu b)]\nonumber\\
\simeq
{1\over \lambda q_{\perp}}\left(\mu\over q_{\perp}\right)^{i\beta N}
\int^{r q_{\perp}}_0 dzJ_1(z)z^{i\beta N}.
\label{eq:T1} 
\eea
The contribution to $I(q_{\perp})$ non-vanishing with growing $N$
comes from $z\sim z_0=\beta N$, where  
$z_0$ is the point of stationary phase. Hence, the requirement
$q_{\perp}\gg \mu \beta N$. 
The width, $\delta z\simeq \sqrt{\beta N}$, of the region, 
the major contribution to (\ref{eq:T1}) comes from, is much smaller
than the interval of integration in (\ref{eq:T1}) which is
$\Delta z \gg \beta N$.
Then,
 for ${\beta N}\gg 1$ one has
\beq                                             
I(q_{\perp})\simeq
{1\over \lambda q_{\perp}}\exp\left[-i\beta N\log{q_{\perp}\over 2\mu}
+2i\varphi\right],
\label{eq:T4} 
\endeq
where
\beq 
\exp[2i\varphi]={\Gamma(1+i\beta N/2)\over \Gamma(1-i\beta N/2)}.
\label{eq:CPHASE}
\eeq
Thus, for    $q_{\perp}\gg \mu\beta N$ one finds no attenuation effect.   
However,  at large $q_{\perp}$  the structure factor of crystal (\ref{eq:SL})
 enters the game.
The resonance  condition (\ref{eq:KAPRES})
implies  fine tuning of phases
of  scattered waves. If $q_{\perp}$ becomes large the phases get out
of tune easily. At large projectile momentum $p$ the structure factor 
(\ref{eq:SL})
allows variation of $q^2_{\perp}$  within a rather wide band
which is however $\propto N^{-1}$.
  In the neighbourhood of the resonance
\beq
S(q_{\perp})\simeq N[1-B^2q^4_{\perp}],
\label{eq:SLS}
\eeq
 where $B=aN/4\sqrt{6}p$.  Hence, the excitation  cross section 
$\sigma=\int dq_{\perp}^2 |T({\bf q}_{\perp})|^2$ 
which  varies as 
\beq
\sigma\propto N^2\log(N_c/N),
\label{eq:SIGMAN2}
\eeq
 where 
\beq
N_c\ll N_n=\left[4\sqrt{6} p_n\over a\mu^2\beta^2\right]^{1/3}
\label{eq:NN}
\eeq
and  $p_n=aM\Delta E/2\pi n$. 
For example, for  the $W$ crystal $N_3\simeq 30$ and for the 
diamond target crystal  one has
 $N_3\simeq 300$.  With certain reservations about the effect of lattice 
 thermal vibrations, one can conclude
 that  not only the $q_{\perp}$-dependence
of the coherent transition amplitude differs
dramatically from the early predictions \cite{PROP2,FUSINA}, but
the effect  of the coherent  enhancement is much weaker than that 
predicted in \cite{PROP1,PROP2,PROP3}.

Notice, that  for $N\ll N_n$ one has
\beq
{p\over q^2_{\perp}}\gg L 
\label{eq:SPA}
\eeq
  and the
 straight paths approximation which we rely upon in our analysis
 holds true. 

We conclude that in quantitative analysis
of high-energy particle-crystal interactions due allowance must be made to
 the multiple scattering effects which 
 dramatically change the pattern
of  the coherent Coulomb excitation compared to widely used approaches
based  on the Born approximation.

{\bf Acknowledgements:} This paper has been inspired by discussions with
L.B.Okun.  Thanks are due to N.N. Nikolaev and B.G. Zakharov for
useful comments. The author thanks J. Speth and FZ-Juelich for hospitality and
DFG (grant 436 RUS 17/119/01) for support. Partial support from INTAS
(grant 97-30494) is gratefully acknowledged.

\newpage\


\begin{thebibliography}{99}
\bibitem{PROP1}
V.V. Okorokov and S.V. Proshin, {\sl Investigation of the coherent 
excitation of the relativistic nuclei in a crystal},
  Moscow, ITEP-13-1980 

\bibitem{PROP2}
Yu.L. Pivovarov, A.A. Shirokov and S.A. Vorobev,   
Nucl. Phys. A509 (1990) 800 ;\\
Yu.L. Pivovarov and A.A. Shirokov, Sov. J. Nucl. Phys. 37 (1983) 653.

\bibitem{FUSINA}
R. Fusina and J.C. Kimball,  Nucl. Instrum. Meth. B33 (1988) 77

\bibitem{PROP3}
V.V. Okorokov, Yu.L. Pivovarov, A.A. Shirokov and S.A. Vorobev,
{\sl Proposal of experiment on coherent excitation of relativistic nuclei in
 crystals},
 Moscow,  ITEP-90-49, Fermilab Library.

\bibitem{GLAU}
R.J. Glauber, in Lectures in Theoretical Physics, edited by W.E. Brittin
et al., Interscience Publishers, Inc., New York, vol.1, p. 315, 1959.

\bibitem{GRIB}
V.N. Gribov, Sov. Phys. JETP 29 (1969) 483; 30 (1970) 709.

\bibitem{KALASH}
N.P.Kalashnikov, E.A.Koptelov and M.I.Ryazanov, Sov. Phys. JETP 63 (1972) 1107;\\
N.P.Kalashnikov and V.D.Mur, Sov. J. Nucl. Phys. 16 (1973) 613.

\bibitem{OKOR} 
V.V. Okorokov, JETP Lett. {\bf 2} (1965) 111; Sov. J. Nucl. Phys. {\bf 2} (1966) 719

\bibitem{EXP}
V.V. Okorokov, D.L Tolchenkov, Yu.P. Cheblukov
et al., JETP Lett 16 (1972); Phys.
Lett. A43 (1973) 485;\\
M.J.Gaillard, J.C.Poizat, J.Remillieux, and M.L.Gaillard, Phys. Lett.
 A45 (1973) 306;\\
H.G. Berry, D.S. Gemmel, R.E. Holland et al., Phys. Lett. A49 (1975) 123;\\
M. Mannami, H. Kudo, M. Matsushita and K. Ishii, Phys. Lett. A64 (1977) 136;\\
S. Datz, C.M. Moak, O.H. Crawford et al., Phys. Rev. Lett. 40 (1987) 843;\\
C.M. Moak, S. Datz, O.H. Crawford et al., Phys. Rev. A19 (1979) 977;\\
F. Fujimoto, Nucl. Instr. Methods B40/41 (1989) 165;\\
Y. Iwata, K. Komaki, Y. Yamazaki et al., Nucl. Instr. Methods, 48 (1990) 163.

\bibitem{GEM}
D.S. Gemmel, Rev.  Mod. Phys. 46 (1974) 129.

\bibitem{Ajzen}
F. Ajzenberg-Selove, Nucl. Phys A190 (1972) 1.

\bibitem{JETPLET}
V.R. Zoller, JETP Lett. 64 (1996) 788.
\end{thebibliography}
\end{document}